# Deep Delay Loop Reservoir Computing for Specific Emitter Identification

Silvija Kokalj-Filipovic, Paul Toliver, William Johnson
Perspecta Labs Inc
Red Bank, New Jersey
skfilipovic@perspectalabs.com

Raymond R. Hoare II, Joseph J. Jezak
Concurrent EDA, LLC
Pittsburgh, Pennsylvania
rayhoare@concurrenteda.com

*Abstract*— Current AI systems at the tactical edge lack the computational resources to support in-situ training and inference for situational awareness, and it is not always practical to leverage backhaul resources due to security, bandwidth, and mission latency requirements. We propose a solution through Deep delay Loop Reservoir Computing (DLR), a processing architecture supporting general machine learning algorithms on compact mobile devices by leveraging delay-loop (DL) reservoir computing in combination with innovative photonic hardware exploiting the inherent speed, and spatial, temporal and wavelength-based processing diversity of signals in the optical domain. DLR delivers reductions in form factor, hardware complexity, power consumption and latency, compared to State-of-the-Art . DLR can be implemented with a single photonic DL and a few electro-optical components. In certain cases multiple DL layers increase learning capacity of the DLR with no added latency. We demonstrate the advantages of DLR on the application of RF Specific Emitter Identification.

I. INTRODUCTION

To maintain military advantage the warfighter at the tactical edge must leverage advancements in artificial intelligence (AI). Current systems lack the computational resources to support in-situ training and inference for situational awareness, and it is not always practical to leverage backhaul resources due to security, bandwidth, and mission latency requirements. We propose a solution through Deep delay Loop Reservoir Computing (DLR), our novel AI processing architecture that supports general retrainable machine learning (ML) solutions on compact mobile devices by leveraging delay-loop reservoir computing (RC) in combination with innovative photonic hardware that exploits the inherent speed and multi-dimensional (spatial, temporal and wavelength-based) processing diversity of signals in the optical domain. Reservoir computing is a bio-inspired approach especially suited for processing time-dependent information in a computationally efficient way [1], [2], [3], [4], [5], [6], [7]. The RC in ML solutions conditions the input features towards linear separability of different classes, upon which any ML algorithm can be trained more efficiently. While leveraging the RC concept, DLR delivers significant reductions in form factor, hardware complexity, and power consumption, providing real-time learning latency that is several orders of magnitude smaller than the State-of-the-Art (SOA) baseline classifiers. For the baseline we utilize 2 neural network models: recurrent (RNN) and residual (ResNet).

In its simplest configuration, DLR can be compactly implemented with a single photonic delay loop and a few electro-optical components, achieving 1000x hardware reduction factor without performance loss compared to the spatial (RNN) implementation. We also discuss how in certain cases multiple delay-loop layers may increase learning capacity of the DLR with no added latency. Similar ideas of multiple coupled loops have appeared recently [8]. DLR combines input transformations and Wavelength Division Multiplexing (WDM) to process multiple inputs in parallel while preserving salient information, which may achieve transformation coding gain and WDM gain by a factor of ten, depending on the application. DLR supports a range of applications, by fine tuning its parameters to achieve low SWaP, high accuracy and required latency. We demonstrate the advantages of the proposed solution on the application of Specific Emitter Identification (SEI), which uses wideband RF samples of 20 different Wi-Fi devices. We provide figures of merit (FOMs) based on algorithmic simulations, the FPGA emulation of the loop and an optical physical layer simulation.

The organization of this paper is as follows: our algorithmic solutions based on the delay-loop concept are described in Section II, based on the SEI dataset (also described here); Section III describes photonic loop implementation; Section IV presents FOMs for the current SEI DLR implementation and concludes.

II. DLR ALGORITHM

*Fig. 1* illustrates the flow of information in the SEI system based on DLR, which has been trained using datasets extracted from captured sequences of RF complex-valued digital samples. Note that real and imaginary parts of digitized RF samples are being referred to as I and Q samples, which stands for *in-phase and quadrature*. Fig. 1 includes the ingest of over-the-air (OTA) RF data samples in order to illustrate the online retraining (retuning) of such a system and the simultaneous use of the trained DLR for inference, i.e., for the real-time specific emitter identification. The (offline) training of such a system is the same - practically contained in the part behind the black circle in Fig. 1 (where data augmentation may be applied for training). In the offline training, the data is coming from a dataset stored as a file.

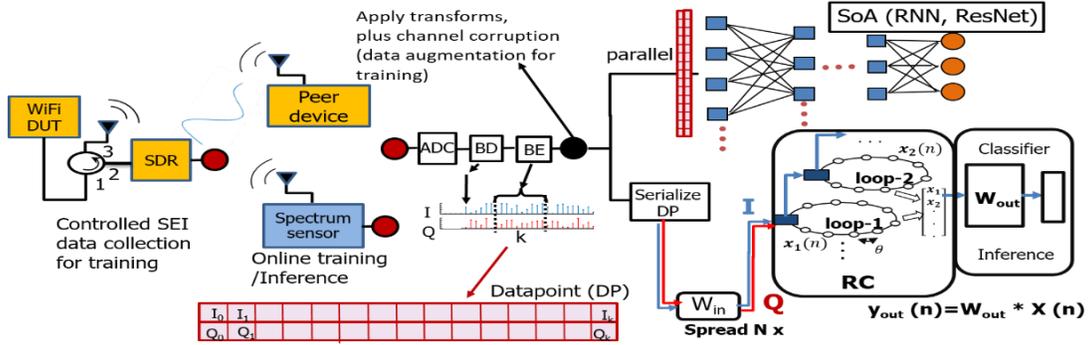

Fig. 1: System architecture and information flow for the DLR used for Specific Emitter Identification (SEI). DUT is Device Under Test; SDR is Software Defined Radio; ADC is Analog-to-Digital Converter; BD is Burst Detector, and BE is Burst Extractor; SoA is State-of-the-Art.

For the illustration of the in-situ online case in Fig. 1, we utilize a Wi-Fi device whose transmission to its peer is captured by the spectrum sensor (usually SDR-based), digitized, and structured into the datapoints comprising k contiguous RF samples, where k is a design parameter.

## A. SEI DATASET

The dataset to train the SEI detector contains 20 classes, corresponding to 20 distinct Wi-Fi devices. To build this dataset we captured the emissions of commercial Wi-Fi devices as they were sending beacons to an access point while using the same spoofed MAC address. We used USRP X310 with a UBX RF daughterboard using a sample rate of 100 MHz centered in the middle of the 2.4 GHz ISM band. Our dataset is carefully prepared to not contain personally identifiable information (PII). Datapoints are created by extracting bursts of $k = 1024$ complex-valued (I/Q) samples from the captured time-series by detecting the energy at the start of a transmission and including the first $k$ samples starting by 500 samples before the detected rising edge. Sub-bursts of 256 complex samples are then extracted to replace original datapoints of 1024-samples after analyzing where in the datapoint the salient information is located (see *Fig. 2* where the yellow surface indicates combinations of start and end indices of the sub-burst that result in highest SEI accuracy among 4 specific Wi-Fi emitters).

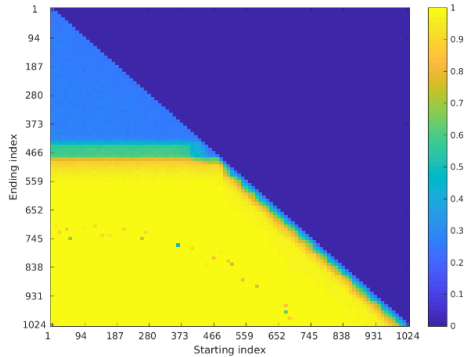

Fig. 2: Saliency map within the original data point of 1024 samples (x axis: burst's start index, y axis: burst's end index)

*Fig. 3* depicts the dataset via three scatterplots of datapoints used for training (100 per class, hence 2000 presented here), where a datapoint is color coded according to its class (1-20). Plot A shows the clean dataset (as captured), where the 256-long datapoints have not been normalized. Note that datapoints are created by using magnitudes of the captured burst of 256 complex samples.

Plots B and C are normalized, i.e., each datapoint is divided by its largest sample (out of 256 here). Datasets in plots B and C are both corrupted by adding random jitter to the carrier frequency (+/- 50kHz), and different noise levels. The corrupted dataset was used to test the performance of the algorithms (both baseline and DLR) on the more realistic SEI data. Note that normalization of the data lowers the accuracy for all compared algorithms. Most of the graphs presented in this paper are obtained for the unnormalized datapoints (more precisely: normalized by a single value - max over all devices).

However, our comparisons based on the graphs and FOMs are still valid as the accuracy decreases with normalization equally for all classifiers. When per-datapoint normalization is used the accuracy figures decrease by ~5% to ~10% depending on whether the data is clean or not. Note that with unnormalized data the power-based differences between classes are discernable even visually (*Fig. 3* A), but they ought to be eliminated since power is not a salient SEI feature, being too dependent on the propagation channel.

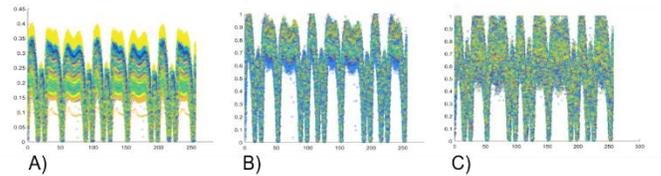

Fig. 3: A) clean unnormalized dataset, B) SNR = 20-30dB + carrier jitter, C) SNR = 10-20dB + carrier jitter

## B. DLR ARCHITECTURE

RC is desirable compared to DNNs since the RC neurons, which have random spatial feedback connections and weights, do not need to be trained. A properly parameterized RC projects the data input into a higher-dimensional state space to achieve better linear separability between data classes for easier classification. Linearly separating datapoints lowers the computational complexity of the training algorithm that ingests such datapoints. The same is true for the delay-loop implementation. This is the feature that is leveraged in the proposed architecture where the training occurs in the read-out block at the output stage of the RC with the stacked states $(x)$ of the RC loops being matched to the labeled ground truth.

Hardware implementation of traditional spatial RC is still an obstacle towards general deployment as an $N$-node reservoir would require $N$ physical neurons. This would cancel out the low computational complexity of the training algorithm (the read-out block in our architecture). Recent work [9] [1] [10] in single loop implementation of a delay feedback RC suggests replacing traditional RC with two functions: i) a photonic delay loop terminating at a nonlinear node, and ii) up-sampling of the input signal and masking with random weights (input mask). Equivalence between delay-loop RC and RNN had been established through the analytic framework of delay differential equation (DDE) [9]. By using the delay loop, we aimed to achieve linear separability without the hardware complexity of a traditional RNN layer (or a spatial RC), as temporal virtual nodes replace the spatial nodes of the traditional RC.

We tested the above assumptions via software implementations of the DLR delay loop algorithm and by performing optical physical layer simulations. We next explain the DLR loop algorithm, and then show that the optimized coupling between the virtual loop nodes, defined by the algorithm and tested via its software implementations (including FPGA), resulted in high SEI accuracy, matching that of the baseline DL models.

Fig. 1 shows that for both the training and the inference, a datapoint $s_i$ of the $k$ samples must be serialized, and then each sample is spread using a random spread sequence $W_{in} = m(t)$, where $m(t)$ contains $N$ chips sampled from $U(-1,1)$. We refer to each of the $N$ elements of the spread sample $J$ as a *chip*. Note that the sample here is not the complex valued I/Q sample, but rather it is transformed into a real number. In the rest of this paper, we will assume that the $i^{th}$ sample is a magnitude of the $i^{th}$ I/Q sample, i.e., $s_i = \sqrt{I_i^2 + Q_i^2}$.

Fig. 4 presents the pseudo code for the loop emulation in which each of the chips $J(n), n = 1, \cdots, N$ of the spread sample $J$ is put through the nonlinearity NL as a linear combination with the oldest virtual element $x(n)$ of the reservoir sate, and then convolved with filter $h(t)$. This is an interpretation of equations (12-17) in [9], where the virtual node k is affected in the following way:

$$x_k(n) = x_k + \int_{\sigma_k - \tau_D}^{\sigma_k} h(\sigma_k - \sigma) f_{NL}[\eta x_\sigma(n-1) + vJ(n-1)]d\sigma. \quad (1)$$

This convolution allows for the coupling of neighboring virtual nodes throughout the datapoint, capturing its recurrent patterns in a way similar to a convolutional network. We parameterized

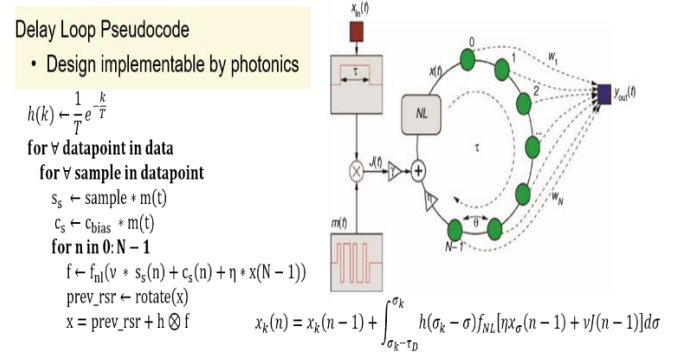

Fig. 4: Sequential emulation of the delay loop

$h(t)$ with the constant $T = 1$.

Our initial experiments began with a dataset of only 4 distinct Wi-Fi devices. Later, we expanded to a larger dataset of 20 devices, requiring adjustment to certain parameters of the sequentialized loop implementation (as in Fig. 4). Surprisingly, a larger number of classes only influenced the reservoir size needed to achieve the baseline accuracy, while the loop gain parameters $v$ and $\eta$ were not affected. However, these parameters required adjustment as the non-linear function used in the loop was changed, which we did for practical reasons. We replaced $f_{NL}(\cdot) = tanh(\cdot)$ with a $\sin^2(\cdot)$ function as the latter is easier to implement in photonics. The parameters $v$ and $\eta$ are also affected by data normalization, which will become the norm in the future. For now we used different parameters for normalized and normalized data.

Note that normalization is a sensitive issue and must be applied carefully. For example if we chose to exercise WDM by feeding I and Q samples into two separate wavelengths, and normalize I and Q vectors independently (by dividing each sample by the max in the I and Q vector, respectively), we will corrupt the phase of the waveform depicted by a given complex datapoint. Hence, normalization should be specific to the applied input data transform. In addition, in the above example the spreading sequences applied to I and Q vectors must be the same. We have tested DLR with separately processed I and Q, but the accuracy seems to be the same as with magnitudes. With WDM this separate processing comes at minimal cost, hence it deserves further investigation and optimization.

Finally, note that the theory of the RC says that linear separability can be achieved by projecting the datapoint into a higher-dimensional space $X \in \mathbb{R}^N$, which for us means $N > L$. With the initial 4-device dataset, utilizing M=200 modified datapoints of L=256 samples per class and reservoir size $N = 400$, the algorithm presented in Fig. 4 resulted in an accuracy above 99%. With the 20-device dataset, the best normalized result of 90% was achieved with the reservoir size $N = 600$ and the same number of samples per class.

**Split Loop:** To lower the DLR complexity by decreasing $N$ further, we applied what we refer to as 'split loop'. Because of the short memory of the filter $h(t)$, we conjectured that the recurrent patterns captured by the reservoir will not be affected if we split the datapoint in half and process each half using two loops in parallel. This allowed us to decrease the reservoir size to any $N > L/2$, and we used $N = 200$ (or even 150) while maintaining the same accuracy. Lower $N$ may also help in reducing the overall footprint of an integrated photonic implementation. The reservoir state of each datapoint is the mean of the final states of the two half-loops.

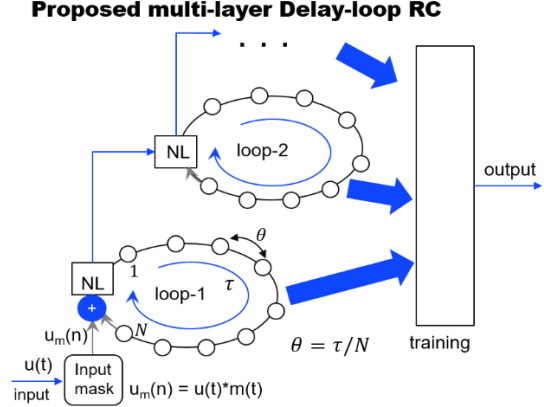

Figure 5: Multiple loops where the output of the 1st loop is shifted at the spreading rate to enter the second loop causing additional *delay of* $\theta = \tau/N$

**Multiple Loops:** Our design idea of a deep DLR composed of multiple loops is based on the reasoning that stacked loops achieve higher flexibility of adjusting the short term temporal memory of the reservoir to provide higher learning capacity for more complex signals. We have partially tested two delay loops. While two loops achieved increased prediction accuracy for Mackey-Glass time series, we have not seen a benefit on the SEI data so far. The loops are stacked sequentially such that the 2$^{nd}$ loop has a delay of $\theta = \tau_1/N$. When the loops are of the same size, we see a decrease in the accuracy of about 15%. Surprisingly, when the second loop is smaller the accuracy decreases by less than 10%. We used the same filter in both loop layers. It remains to be seen if the accuracy matching or exceeding the single loop can be recovered by fine-tuning the filters. With split loops, the two-loop architecture is implemented such that each split loop feeds into one of the split loops in the following layer.

**Read-out and training:** For training the SEI classifier, we propose to deploy Ridge Regression (RR) described by:

$$W = \underset{W_{out}}{argmin} \left\{ \sum_{j=1}^{B} \|y_j - W_{out} * x\|_2^2 + \lambda \|W_{out} x\|_2^2 \right\} \quad (2),$$

where $B$ is number of training datapoints, $y_j$ is the device-label corresponding to the $j^{th}$ training datapoint, $W_{out}$ is output weights, and $\lambda$ is the regularization factor. Equation (2) can also be written in the closed form $W = (X'X + \lambda I_N)^{-1} (X'Y_{out})$, where $X$ is the matrix of the stacked state vectors $x$ for all training datapoits, and $Y_{out}$ is the matrix of matching labels in *one-hot* encoding. Note that the inference is performed with trained weights in the following way: a datapoint is sensed and extracted OTA, fed into the RC (one or multiple loops), and the final state vector $x$ is then multiplied by $W$ to give the device label for the datapoint: $y_{out} = xW$. We have also evaluated a shallow fully connected neural network (FCNN) and a shallow LSTM NN to replace RR. We needed the NN to be shallow in order to avoid an increase in training complexity, while potentially allowing for transfer learning and distributed learning in the future. However, these NN architectures failed to outperform the current method.

Note that training the weight coefficients using the closed form RR does not require multiple iterations of the algorithm. Hence, both its accuracy and its complexity are deterministic while the NN approaches are stochastic, and their accuracy fluctuates even after the algorithm has converged. The slight random variation in the DLR approach with RR is solely caused by the reservoir computer (loop) due to the random spreading sequence and random training batches. Note that for the production training, the spreading sequence $m(t)$ can be optimized and fixed.

With the FCNN, we used the same state vectors $X$ from the output of the loop and fed it into both the *FCNN* and the *RR*. When the FCNN failed to outperform the current method, we next used a recurrent network (LSTM) of lower capacity than the baseline (almost twice). This network did not perform well with the original datapoints but when fed with the state vectors it approached the baseline accuracy. However, the complexity of such a NN is still much higher than the RR approach.

### III PHOTONIC LOOP IMPLEMENTATION

Our proposed photonic implementation of the RC architecture, shown in Fig. *6*, combines the strengths of photonics and electronics while providing a scalable path to future chip-scale implementations. Here, the analog input $u_{in}(t)$ is sampled for duration $\tau$ and upconverted to an optical frequency. A random input mask $W_{in}(t)$ with a sample rate $\theta$ is then applied. This signal is sent into the reservoir layer that comprises optical-to-electrical conversion, followed by a $sin^2(\cdot)$ nonlinearity obtained from an intensity modulator with a response time $TR_1 \sim 5\theta$. Assuming a loop delay time of $\tau = N\theta$, the reservoir loop supports $N$ virtual nodes, which are continuously updated on each round trip.

By leveraging WDM, a single delay line allows the number of virtual nodes to be scaled by a factor $P$. A dynamic gain equalizer (DGE) enables round trip loss to be equalized across the wavelength channels. Such a DGE is implementable using micro-ring resonators [11], making it amenable to photonic integration. The reservoir may be extended to multiple RC loops, each of which may have different response and/or round-trip times. The reservoir readout is performed here using a serialized output layer, which is achieved by tapping each of the reservoir layers.

The output signals from different reservoirs are time multiplexed using appropriate delays and a switch. Bipolar output weights, $W_{out}(t)$, are applied at sample rate $\theta$ using a dual-output E/O modulator in combination with a balanced photodiode [10]. Finally, a low pass filter is used to compute a dot product sum at the final output.

To ensure input, output, and nonlinear operations are independent, associated elements are implemented on a per wavelength basis, as shown in Fig. *6*.

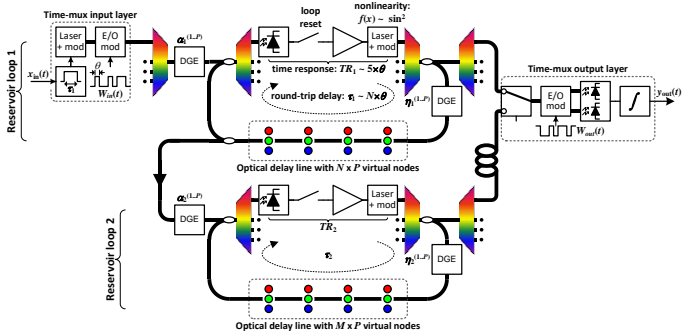

Fig. 6: Photonic reservoir computing architecture employing WDM. E/O: electro-optic, mod: modulator, DGE: dynamic gain equalizer

The loop-gain $G$ is related to the number of times $K$ the signal traverses the loop and the one-pass gain $\eta$ by $G = \eta^K$. In steady-state, the Barkhausen criteria to avoid oscillation is $G<1$. Our simulations show that this can be relaxed by periodically opening the loop so oscillations do not build, which only requires $G_{ave} = \frac{1}{K+1}\sum_{k=1}^{K} \alpha^k < 1$. This includes the effect of gain variation due to optical coupling and the finite electronic bandwidth. In this case we do not have to regulate the loop-gain in real time to achieve loop gain with $\alpha = 1^-$ and $K = 1000$.

Our initial single-loop hardware testbed for analyzing the photonic delay loop-based DLR is is shown in Fig. 7, which uses electrical summation into the delay loop rather than the optical summing shown previously. We have constructed both MATLAB and VPI TransmissionMaker simulation frameworks for comparing the predicted performance and measured results from our testbed. In the testbed, input waveforms are generated offline using the MATLAB DLR model and downloaded to memory of a high-speed arbitrary waveform generator (AWG) before outputting at a 1 GSa/s sample rate. An electrical amplifier with >6 GHz bandwidth is used to scale the amplitude prior to a broadband electrical summing hybrid. An electrical switch is used to reset the delay loop to its initial state prior to injecting the input electrical waveform. Finally, a distributed feedback (DFB) laser and a high-speed (BW > 10 GHz) lithium niobate Mach-Zehnder (MZ) modulator serve as the nonlinear element within the optical delay loop, which is terminated by a detector and filtered to a 200 MHz bandwidth prior to data acquisition (DAQ). We are constructing the initial DLR testbed to support 256 virtual nodes with a 1 ns temporal spacing, resulting an optical fiber length of ~51 meters. We plan to later investigate higher speed sampling and signal injection, with the goal of reducing the fiber delay to a length compatible with photonic integration.

As an additional all-electronic hardware emulation platform, an FPGA implementation of the DLR was also implemented based directly on our MATLAB DLR simulation framework. This code was converted into double-precision C software and optimized to only utilize single-precision operators. To achieve maximum performance from the optimized FPGA implementation, the inner loop was unrolled.

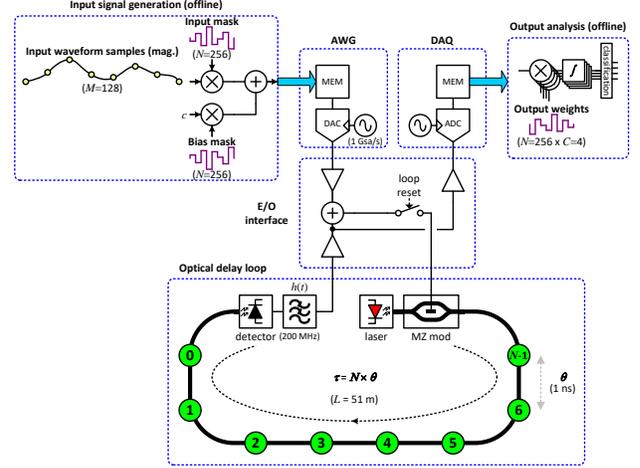

Fig. 7: Hardware testbed for analyzing photonic delay loop-based DLR

From this optimized software implementation in C, the DLR code was then analyzed with Concurrent EDA's C-to-Gates compiler, which generated VHDL including a 256-stage pipeline to match the DLR algorithm's delay. The $sin()$ function was implemented as a lookup table with interpolation between two results to ensure accuracy. These changes helped to ensure that a sample may be processed on every cycle, preventing stalls due to data dependencies and ensuring maximum throughput.

This H/W implementation achieved a frequency of 200MHz, allowing the DLR to iterate every 5ns. Detailed simulation of the VHDL showed a performance of 356us for a reservoir size of 256 and a burst size of 256. The FPGA area was less than 2% of the available resources on the Xilinx RFSoC chip. At this size, it is feasible to implement up to 40 such DLR processors in a single chip or dramatically increase the number of classes.

## IV FIGURES OF MERIT

Table 1 below shows important figures from which we derive FOMs for the training, by applying some very elaborate calculations. Table 1 reflects the latest results utilizing the new dataset with 20 classes. Accuracy figures are based on independent test dataset. $M_{DLR}$ (memory in Table 1) and $C_{DLR}$ (complexity) are determined by the matrix of RC states X, of size $B*N$, whose pseudo-inverse has complexity $\boldsymbol{BN^2}$. We have revised $B$ and $N$ to: $B = 8000$, and $N = 800$. Observe that DLR complexity does not scale with number of classes. Please note that we use 1000 samples per device to train the SOA networks and 400 samples per device for DLR. With almost 800 epochs needed for RNN training to converge, we obtained its total complexity shown in Table 1, which on the current system with 1 GPU resulted in more than 12 hours of training, while the DLR training on the CPU took only one hour. Note that the inference complexity of DLR $C_{DLR}^I$ is now reduced to $C_{DLR}^I = QN^2 = 7.2e6$, more than 1000 times lower than RNN.

*Table 1: Basic Performance Figures*

| Figure \ Model | Accuracy (%) | # of parameters (memory M) | Complexity (C) |
|---|---|---|---|
| ResNet | 95.6 | 274K | C=8.5e12 |
| LSTM RNN | 95.0 | 2.1M | C=6.5e13 |
| DLR | 95.1 | Q*N = 16K | $(N^2 B)=(400*20)^3 = 5.e11$ |

The ADC that is built-in on the RFSOC platform can support a maximum input bandwidth of 2 GHz, which will set an upper limit on the rate at which information can be retrieved from the photonic loop implementation. If the full bandwidth were usable, this would result in a $100\,ns$ per spread-sample processing time in the photonic loop implementation. The processing time per data point $\Delta_d = 26us$. Please compare this to 0.356ms per datapoint that we currently achieved with FPGA. Hence, the RC would incur *training latency* of $\Delta_{RC} = \Delta_d * B \approx 208ms$ during collection of the state matrix X. We add the memory read of X, to calculate the total photonic loop latency for the 20-device dataset as $\Delta'_{DLR} = \Delta_{RC} + 2 * M_{DLR} * \frac{1}{f_{bus}}$. As the bus clock is 256MHz, $\Delta'_{DLR} \approx 208ms$. Total DLR training latency $\Delta_{DLR} = \Delta'_{DLR} + \Delta_{RR}$ where we added a factor due to Ridge regression (not optimized yet, but currently around 500ms at the ARM processor on RFSoC). This is ~ 1s for training versus 1 hour in MATLAB and more than 12 hours for RNN training (x 700).

In the table below, we give FOMs for the training mode, except for the latency reduction factor (LRF) where we also add inference latency since it is more precise at this point. The FPGA *inference latency* per datapoint is $\Delta_d = 356\mu s$ versus the RNN *inference latency* per datapoint, which is $\sim 1700 \mu s$. This can be further optimized. *The revised training FOMs (wrt RNN)* for the 20-device SEI are in the table below, where SR denotes spatial reduction, HCR denotes hardware complexity reduction, and LR is latency reduction:

| |
|---|
| SR FOM  (SRF) = $\frac{M_{RNN}}{M_{DLR}} \approx 200$ |
| HCR FOM  (HCRF) = $\frac{C_{RNN}}{C_{DLR}} \approx 130$ |
| LR FOM (LRF) = $\frac{\Delta_{RNN}}{\Delta_{DLR}} \approx$ ~700 (~5 for inference). |

Note that the above FOMs are temporary as we continue working on optimizing the DLR and introducing new feature transforms, as well as proceeding to implement all its parts in hardware.

ACKNOWLEDGMENT

This research was, in part, funded by the U.S. Government. The views and conclusions contained in this document are those of the authors and should not be interpreted as representing the official policies, either expressed or implied, of the U.S. Government. **DISTRIBUTION STATEMENT A. Approved for public release: distribution unlimited**

REFERENCES

1. *Advances in photonic reservoir computing.* **al., Van der Sande et.** s.l. : Nanophotonics, 6(3), 2017.

2. *Reservoir computing approaches to recurrent neural network training.* **Lukoševičius M, Jaeger H.** 2009, Comput. Sci. Rev., Vol. 3, pp. 127-49.

3. *An experimental unification of reservoir computing methods.* **Verstraeten D, and Schrauwen B, and D'Haene M, and Stroobandt D.** 2007, Neural Networks, Vol. 20, pp. 391-403.

4. *Short term memory in echo state networks. .* **H., Jaeger.** 2001, German National Research Center for Information Technology, Technical Report GMD Report , Vol. 152.

5. *Minimum complexity echo state network. .* **Rodan A, Tino P.** 2011, IEEE Trans Neural Netw , Vol. 22, pp. 131–44.

6. *Information processing using a single dynamical node as complex system.* **Appeltant L, Soriano MC, Van der Sande G, et al.** 468, 2011, Nat Commun , Vol. 2.

7. *Optoelectronic reservoir computing. .* **Paquot Y, Duport F, Smerieri A, et al.** 287, 2012, Sci Rep , Vol. 2.

8. *Coupled nonlinear delay systems as deep convolutional neural networks.* **B. Penkovsky, et al.** s.l. : ArXiv , 2019.

9. *Tutorial: Photonic neural networks in delay systems.* **D. Brunner, B. Penkovsky, B. A. Marquez, M. Jacquot, I. Fischer, and L. Larger.** s.l. : Journal of Applied Physics, 2018.

10. *Fully analogue photonic reservoir computer.* **al., F. Duport et.** s.l. : Scientific Reports 6 , 2016.

11. *Fully programmable ring-resonator-based integrated photonic circuit for phase coherent applications.* **al., A. Agarwal et.** s.l. : Journal of Lightwave Technology 24, 2006.

12. *https://www.xilinx.com/products/boards-and-kits/zcu111.html.* **Xilinx.** 2019.

13. *Continuously active interferometer stabilization and control for time-bin entanglement distribution .* **P. Toliver, et. al.** s.l. : Optics Express 23, 2015.

14. *A Programmable Optical Filter Unit Cell Element for High Resolution RF Signal Processing in Silicon Photonics.* **P. Toliver, et. al.** San Diego, USA : Optical Fiber Communication (OFC), 2010.